\documentclass[11pt, letterpaper, logo, onecolumn, copyright]{main}

\usepackage[authoryear, sort&compress, round]{natbib}

\usepackage[inkscapeformat=png]{svg}

\usepackage[most, breakable, skins]{tcolorbox}

\tcbuselibrary{skins}
\usepackage{lipsum}
\usepackage{tabularx}
\usepackage{afterpage}
\usepackage{booktabs}
\usepackage{xurl} 
\usepackage{subcaption}
\usepackage{makecell}
\usepackage{multirow}
\usepackage{bm}
\usepackage{multicol}
\usepackage{array}
\usepackage{float}
\usepackage{listings, listings-rust}
\usepackage{fontawesome5}
\usepackage{hyperref}
\usepackage{amssymb}
\usepackage{graphicx}
\usepackage[dvipsnames]{xcolor}
\usepackage{cleveref}
\usepackage{longtable}
\usepackage{pdflscape}
\usepackage{adjustbox}
\usepackage{nicematrix}
\usepackage{CJKutf8}
\usepackage{ragged2e}
\usepackage{colortbl}
\usepackage{enumitem}
\usepackage{bbm}
\usepackage{pifont}
\usepackage{hyphenat}
\usepackage{amsmath}
\usepackage{amsthm}  
\usepackage{circledsteps}
\usepackage{diagbox}
\usepackage{bookmark}
\usepackage{textcomp}
\usepackage{fvextra}

\usepackage{algorithm} 
\usepackage{algpseudocode}

\usepackage{wrapfig}
\usepackage{xspace}
\usepackage{tikz}
\usepackage[normalem]{ulem}

\usepackage{pgfplots}
\usepackage{pgfplotstable}
\pgfplotsset{compat=1.18}

\usepackage[most]{tcolorbox}
\definecolor{lightblue}{RGB}{210, 220, 250}

\definecolor{medgray55}{gray}{0.55}
\definecolor{medgray}{gray}{0.7}
\definecolor{litegray}{gray}{0.9}
\definecolor{gblue}{RGB}{210, 227, 252}
\definecolor{gred}{RGB}{250, 210, 207}
\definecolor{gyellow}{RGB}{254, 239, 195}
\definecolor{ggreen}{RGB}{206, 234, 214}
\definecolor{gorange}{RGB}{254, 223, 200}

\definecolor{gblue9}{RGB}{23, 78, 166}
\definecolor{gred9}{RGB}{165, 14, 14}
\definecolor{gyellow9}{RGB}{227, 116, 0}
\definecolor{ggreen9}{RGB}{13, 101, 45}
\definecolor{gorange9}{RGB}{176, 96, 0}

\definecolor{myblue}{rgb}{0,0,1}
\definecolor{myred}{rgb}{1,0,0}
\definecolor{mylightgray}{gray}{0.95}
\definecolor{myCite}{HTML}{1C4587}

\definecolor{highlightblue}{HTML}{185ABC}
\definecolor{cellHighlight}{HTML}{dbefff}
\definecolor{lightgray}{RGB}{211, 211, 211}
\definecolor{lightfont}{gray}{0.3}

\usepackage{minitoc}

\noptcrule


\newcolumntype{L}[1]{>{\raggedright\let\newline\\\arraybackslash\hspace{0pt}}m{#1}}
\newcolumntype{C}[1]{>{\centering}m{#1}}

\newcolumntype{R}[1]{>{\raggedleft\let\newline\\\arraybackslash\hspace{0pt}}m{#1}}

\definecolor{ao}{rgb}{0.0, 0.0, 1.0}

\newcommand\vcent[1]{\vcenter{\hbox{#1}}}

\newcommand\loudspeaker[1][3]{\ensuremath{\vcent{\rule{.6ex}{.6ex}}\kern-.5ex
  \vcent{\scalebox{.6}[1]{\rotatebox[origin=center]{90}{$\blacktriangle$}}}
  \ifnum#1>0\relax\kern.05ex\vcent{\scalebox{.4}{\ttfamily)}}
  \ifnum#1>1\relax\kern-.4ex\vcent{\scalebox{.56}{\ttfamily)}}
  \ifnum#1>2\relax\kern-.55ex\vcent{\scalebox{.7}{\ttfamily)}}
  \fi\fi\fi}
}

\makeatletter
\renewcommand\subparagraph{
 \@startsection {subparagraph}{5}{\z@ }{3.25ex \@plus 1ex
 \@minus .2ex}{-1em}{\normalfont \normalsize \bfseries }}
\makeatother

\let\cite\citep
\hypersetup{
  citecolor = myCite,  
  linkcolor = myCite,   
  urlcolor  = myCite
}
\newcommand{\myheaderbreak}{\\}
\title{
Muse: Towards Reproducible Long-Form Song Generation with Fine-Grained Style Control
}

\author{
  Changhao Jiang\textsuperscript{\rm *\dag},
  Jiahao Chen\textsuperscript{\rm *},
  Zhenghao Xiang\textsuperscript{\rm *},\\
  \bfseries
  Zhixiong Yang\textsuperscript{\rm *},
  Hanchen Wang\textsuperscript{\rm *},
  Jiabao Zhuang\textsuperscript{\rm *},\\
  Xinmeng Che\textsuperscript{\rm },
  Jiajun Sun\textsuperscript{\rm },
  Hui Li\textsuperscript{\rm },
  Yifei Cao\textsuperscript{\rm },
  Shihan Dou\textsuperscript{\rm },\\
  Ming Zhang\textsuperscript{\rm },
  Junjie Ye\textsuperscript{\rm },
  Tao Ji\textsuperscript{\rm \dag},
  Tao Gui\textsuperscript{\rm \dag},
  Qi Zhang\textsuperscript{\rm },
  Xuanjing Huang\textsuperscript{\rm }\\
  \vspace{0.3cm} 
  \normalsize 
  Fudan NLP Group \\
  \texttt{\normalsize chjiang25@m.fudan.edu.cn, \{taoji,tgui\}@fudan.edu.cn}
}

\vspace{-50pt}

\begin{abstract}
Recent commercial systems such as Suno demonstrate strong capabilities in long-form song generation, while academic research remains largely non-reproducible due to the lack of publicly available training data, hindering fair comparison and progress.
To this end, we release a fully open-source system for long-form song generation with fine-grained style conditioning, including a licensed synthetic dataset, training and evaluation pipelines, and Muse, an easy-to-deploy song generation model.
The dataset consists of 116k fully licensed synthetic songs with automatically generated lyrics and style descriptions paired with audio synthesized by SunoV5.
We train Muse via single-stage supervised finetuning of a Qwen-based language model extended with discrete audio tokens using MuCodec, without task-specific losses, auxiliary objectives, or additional architectural components.
Our evaluations find that although Muse is trained with a modest data scale and model size, it achieves competitive performance on phoneme error rate, text--music style similarity, and audio aesthetic quality, while enabling controllable segment-level generation across different musical structures.
All data, model weights, and training and evaluation pipelines will be publicly released, paving the way for continued progress in controllable long-form song generation research.
The project repository is available at https://github.com/yuhui1038/Muse.
\end{abstract}

\begin{document}
\doparttoc
\faketableofcontents
\begingroup
  \renewcommand\thefootnote{}
  \footnote{\hspace{-1.8em}\textsuperscript{*}Equal Contribution.\\
            \textsuperscript{\dag}Corresponding Author.}
\endgroup
\vspace{-30pt}
\maketitle
\renewcommand{\myheaderbreak}{ } 
\vspace{-15pt}

\begin{figure*}[tbh]
\begin{center}
\includegraphics[width=\textwidth]{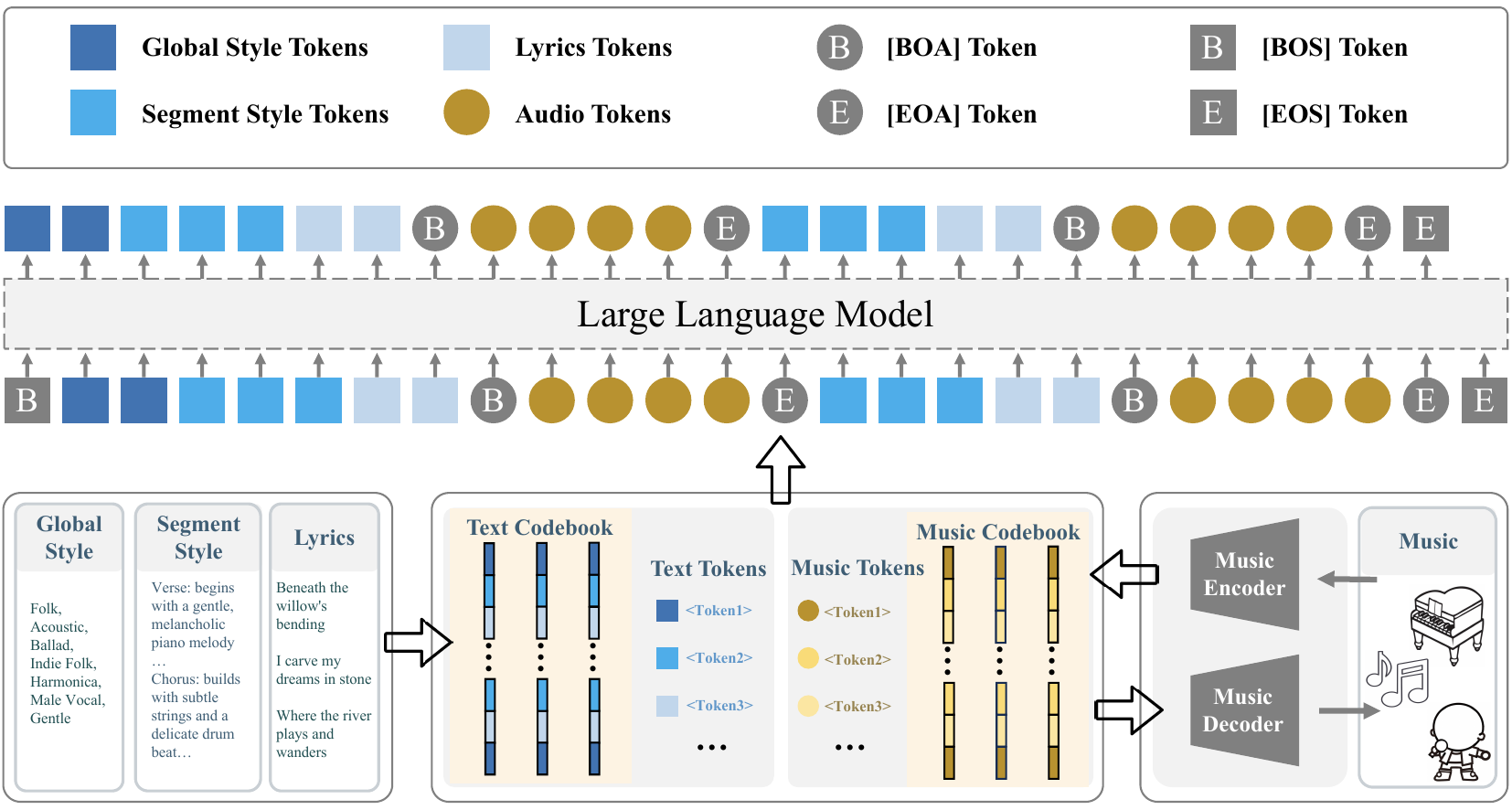}
\caption{Overview of Muse. The model operates on conversational, segment-structured inputs including global style labels, segment-level style descriptions, and lyrics. Text inputs are tokenized with a standard language tokenizer, and audio waveforms are encoded into discrete tokens via a neural audio codec. Text and audio tokens are unified into a single autoregressive sequence, enabling long-form song generation with segment-level style conditioning. [BOA]/[EOA] tokens mark audio boundaries, while [BOS]/[EOS] indicate full-sequence boundaries.}
\label{intro}
\end{center}
\vskip -0.25in
\end{figure*}

\section{Introduction}
Long-form song generation aims to produce complete songs that integrate vocals, lyrics, and musical structure over several minutes of audio. Compared to short-form music generation, this task requires modeling long-range temporal coherence, alignment between lyrics and vocals, and consistency across different structural segments of a song. Recent systems have demonstrated impressive generation quality, suggesting that end-to-end song generation is increasingly feasible \citep{suno2024}.

Despite these advances, academic research in long-form song generation remains largely non-reproducible. Most existing systems capable of generating full songs, including Diffrhythm \citep{ning2025diffrhythm}, LeVo \citep{DBLP:journals/corr/abs-2506-07520}, YuE \citep{DBLP:journals/corr/abs-2503-08638}, and ACE-Step \citep{DBLP:journals/corr/abs-2506-00045}, do not release their training data, while commercial systems such as Suno \citep{suno2024} expose only proprietary APIs. As a result, reported performance is difficult to verify, comparisons across methods are often unfair, and progress in this area is hard to measure or build upon.

In this work, we introduce Muse, a fully open-source model for long-form song generation with fine-grained style conditioning (see Figure~\ref{intro}). The key challenge in releasing song-generation datasets lies in copyright constraints, which severely limit the availability of licensed music suitable for open research \citep{DBLP:journals/corr/abs-2005-00341}. To overcome this barrier, we construct a large-scale dataset of 116k fully licensed synthetic songs by conditioning SunoV5 \citep{suno2024} on GPT-generated prompts \citep{openai2025gpt5systemcard} consisting of lyrics and global style labels. Each song includes structured, time-aligned lyrics returned by SunoV5, together with explicit global style annotations and segment-level style labels automatically derived using an audio–language model, enabling hierarchical style control during generation.

Muse is trained using a simple and reproducible pipeline. We adopt a Qwen-based language model \citep{DBLP:journals/corr/abs-2505-09388} extended with discrete audio tokens obtained via MuCodec \citep{xu2024mucodec}, and perform a single-stage supervised finetuning without task-specific losses or auxiliary components. Despite this minimal design and modest data scale, Muse achieves near state-of-the-art performance across multiple objective metrics, including phoneme error rate, text–music alignment, and audio aesthetic quality. Moreover, Muse enables fine-grained segment-level style control across different structural components of a song, a capability that has not been previously available in open-source song generation models.
Overall, our main contributions are as follows:
\begin{enumerate}
    \item We release a fully open-source system for long-form song generation with fine-grained style control, including Muse, an easy-to-deploy song generation model, along with complete training and evaluation pipelines.
    \item We construct and release a large-scale, fully licensed synthetic dataset of complete songs with structured lyrics and segment-level style annotations, addressing a key reproducibility bottleneck in music generation research.
    \item We demonstrate that a simple single-stage supervised finetuning approach can achieve competitive performance while supporting fine-grained segment-level style control, without relying on specialized architectures or task-specific losses.
\end{enumerate}

\section{Related Work}

\paragraph{Long-Form Song Generation.}
Early work demonstrated the feasibility of minutes-long music generation using codec-based autoregressive modeling, most notably Jukebox, which conditioned on artist or genre labels with loosely aligned lyrics and proprietary data \citep{DBLP:journals/corr/abs-2005-00341}. Subsequent text-to-music systems such as MusicLM~(\citeyear{DBLP:journals/corr/abs-2301-11325}) and MusicGen~(\citeyear{DBLP:journals/corr/abs-2306-05284}) improved audio fidelity and text adherence through hierarchical or single-stage language modeling, but primarily focused on instrumental or short-form generation.

More recent research targets full-song generation with vocals and accompaniment. Song generation models such as MelodyLM~(\citeyear{DBLP:conf/acl/HongHCWLYZZ24}), SongCreator~(\citeyear{DBLP:conf/nips/Lei0TLLLWK0M24}), YuE~(\citeyear{ning2025diffrhythm}), DiffRhythm~(\citeyear{DBLP:journals/corr/abs-2507-12890}), DiffRhythm+~(\citeyear{diffrhythm2}), DiffRhythm2~(\citeyear{DBLP:journals/corr/abs-2506-07634}), SongBloom~(\citeyear{DBLP:journals/corr/abs-2503-08638}), ACE-Step~(\citeyear{DBLP:journals/corr/abs-2506-00045}), SongGen~(\citeyear{DBLP:conf/icml/LiuDZD0ZCL025}), and LeVo~(\citeyear{DBLP:journals/corr/abs-2506-07520}) explore autoregressive, diffusion-based, or hybrid approaches for generating multi-minute songs. While these systems demonstrate strong generation quality, most rely on undisclosed or proprietary training data. As a result, reproducibility and fair comparison remain limited, and commercial systems such as Suno~(\citeyear{suno2024}), Udio~(\citeyear{udio2025}) and Mureka~(\citeyear{mureka2025}) remain closed benchmarks rather than open research baselines.

\begin{figure*}[t]
\begin{center}
\includegraphics[width=\textwidth]{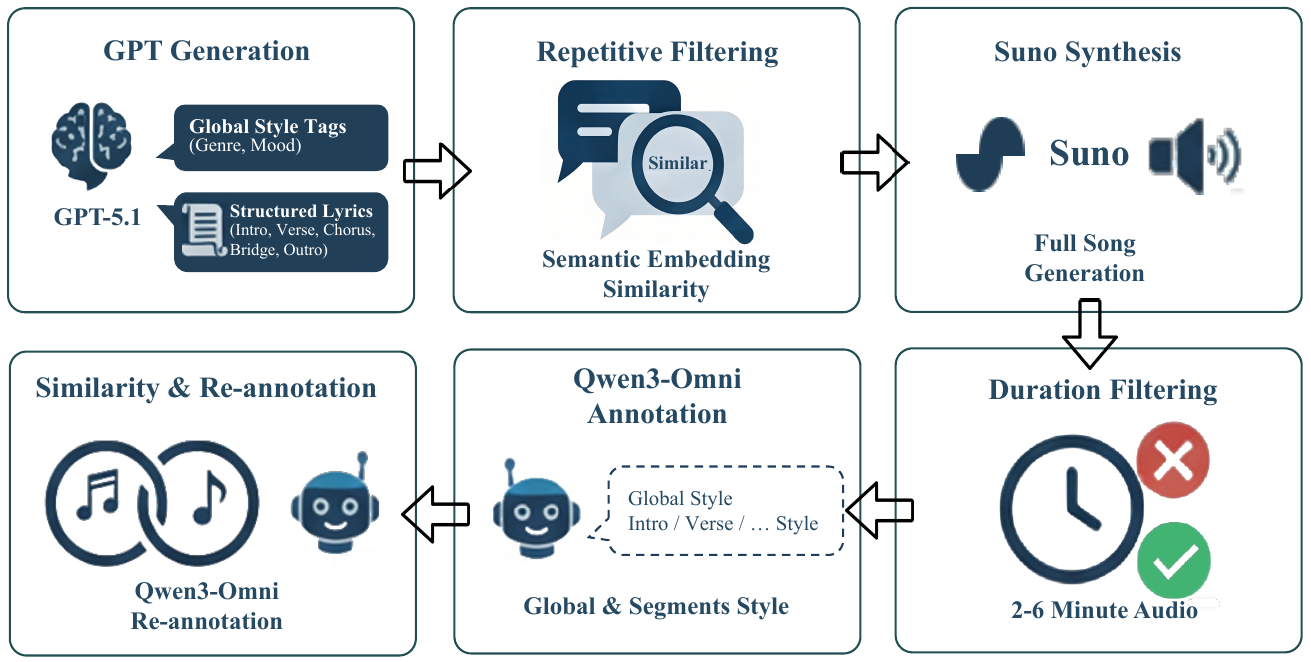}
\caption{Overview of the synthetic song data generation pipeline. GPT-5 mini generates structured prompts with global style labels and segmented lyrics, which are used by SunoV5 to synthesize full-length songs. Qwen3-Omni then produces hierarchical style annotations for each song.}
\label{data_pipeline}
\end{center}
\end{figure*}

\paragraph{Style Control in Song Generation.}

Style control in song generation has largely been implemented as global conditioning via text prompts, such as genre, mood, or artist descriptors. Early systems including Jukebox~(\citeyear{DBLP:journals/corr/abs-2005-00341}), MusicLM~(\citeyear{DBLP:journals/corr/abs-2301-11325}), and MusicGen~(\citeyear{DBLP:journals/corr/abs-2306-05284}) supported high-level stylistic guidance but did not provide explicit control over different parts of a song. Text-to-song models further improved alignment between lyrics, vocals, and accompaniment through architectural constraints or hybrid conditioning signals, as in MelodyLM~(\citeyear{DBLP:conf/acl/HongHCWLYZZ24}) and SongCreator~(\citeyear{DBLP:conf/nips/Lei0TLLLWK0M24}).

Recent systems report broader controllability over musical attributes such as instrumentation or timbre, but these controls are typically applied at the global or track level \citep{DBLP:conf/icml/LiuDZD0ZCL025,DBLP:journals/corr/abs-2506-00045}. More recently, a few works have begun to explore temporally structured or segment-level control, such as TVC-MusicGen and SegTune, though they focus on instrumental music or adopt non-autoregressive, task-specific frameworks \citep{DBLP:conf/interspeech/YangCWZ025,DBLP:journals/corr/abs-2510-18416}. Explicit, user-addressable segment-level style conditioning aligned with song structure remains uncommon in fully open-source settings, and the lack of publicly available datasets with segment annotations further limits reproducible evaluation. Muse addresses this gap by enabling segment-level style control within a fully open and reproducible framework.

\section{Dataset}
\label{sec:dataset}

To enable reproducible research on long-form song generation, we construct a large-scale dataset of fully licensed synthetic songs. Existing full-song generation models are typically trained on proprietary or undisclosed music corpora, making fair comparison and replication difficult. Our dataset addresses this limitation by providing complete, fully licensed songs with structured lyrics and hierarchical style annotations, enabling reproducible research and open redistribution.

\subsection{Song Generation Pipeline}
Each song in our dataset is generated through a fully automated pipeline (see Figure \ref{data_pipeline}). We first use GPT-5 mini \citep{openai2025gpt5systemcard} to generate a textual prompt consisting of two components: (1) a set of global style labels describing high-level musical attributes such as genre, mood, and vocal characteristics, and (2) a complete set of lyrics explicitly structured according to standard song forms, including segments such as Intro, Verse, Chorus, Bridge, and Outro.

These prompts are then provided to SunoV5 \citep{suno2024}, which synthesizes a complete song conditioned on the global style labels and structured lyrics. The output is a single-track audio recording containing both vocals and accompaniment, together with time-aligned lyrics returned by the generation system. All samples correspond to complete, full-length songs, and we retain both the audio and the associated lyrics without separating vocal and instrumental tracks or relying on symbolic representations such as MIDI \citep{midi1996spec}.

This pipeline ensures that each song is associated with explicit textual descriptions at the song level and a predefined lyrical segmentation, while avoiding the use of any copyrighted source audio during data construction.

\subsection{Style Annotation and Refinement}
We annotate each song with both global and segment-level styles to enable fine-grained control and reliable supervision during training.

\paragraph{Global Style Verification.} 
Although SunoV5 is capable of following high-level style instructions, we observe that some generated songs only partially conform to the provided global style labels. To improve annotation reliability, we apply an automatic verification and refinement procedure based on text–music similarity.

For each generated song, we compute the similarity between its global style labels and the corresponding audio using a text–music similarity model, MuQ-MuLan \citep{DBLP:journals/corr/abs-2501-01108}. We observe that approximately 13\% of the songs exhibit a similarity score below 0.25, with an overall average similarity of 0.45, indicating noticeable mismatches between the intended style prompts and the realized audio.

To improve annotation consistency, we re-annotate the global styles of all songs using a large-scale audio–language model, Qwen3-Omni-30B-A3B-Thinking \citep{DBLP:journals/corr/abs-2509-17765}. The model is applied as an automatic audio-to-text annotator conditioned on the generated audio. After re-annotation, the average text–music similarity increases to 0.58, and the proportion of songs with similarity below 0.25 is reduced to 0.03\%. This refinement process substantially improves the alignment between style annotations and musical content, without discarding any generated audio.

\paragraph{Segment-Level Style Annotation.} 
Beyond global style, we further annotate each song with segment-level natural-language style descriptions aligned with its lyrical structure. Given the predefined segmentation of lyrics and the corresponding audio, we employ the same audio–language model to generate detailed textual style descriptions for each structural segment. These descriptions are validated using text–music similarity, and segments with similarity scores below 0.25 are re-annotated to ensure semantic consistency between the audio content and the associated style descriptions.

As a result, each song is associated with both global and segment-level style annotations, enabling hierarchical and fine-grained control during model training and evaluation.

\subsection{Data Filtering and Deduplication}
We apply minimal filtering to preserve data diversity and ensure transparency. Prior to audio synthesis, we perform deduplication on the GPT-generated textual prompts, including lyrics and global style labels, using semantic embedding similarity to remove near-duplicate samples. 

After song generation with SunoV5, we apply a lightweight duration-based filter to the resulting audio. From an initial pool of approximately 118k generated songs, we remove around 2k samples whose durations fall outside the target range of 2–6 minutes. This duration constraint is the only filtering criterion applied to the audio data.

Given the high audio quality of SunoV5 outputs, we do not apply loudness normalization or any other post-processing steps.

\begin{figure}[t]
\begin{center}
\includegraphics[width=0.6\linewidth]{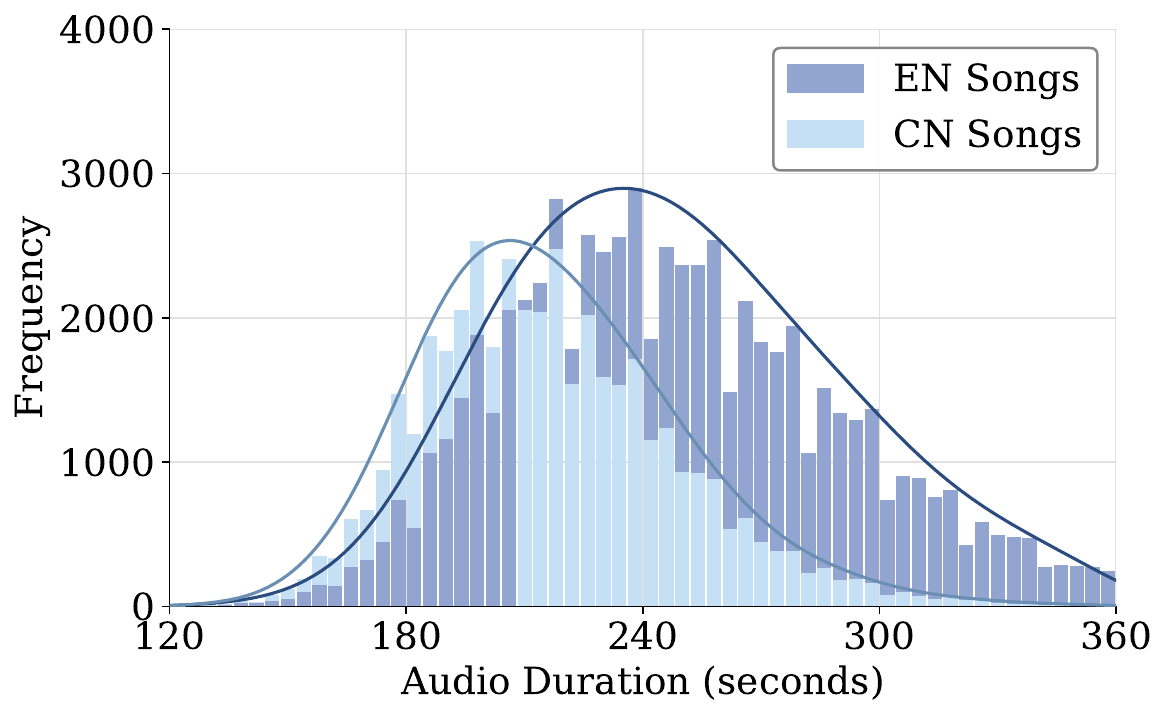}
\caption{Duration distributions of Chinese and English songs in the dataset, showing comparable length profiles across languages.}
\label{fig:duration}
\end{center}
\end{figure}

\subsection{Dataset Statistics}
The final dataset contains 116,489 complete songs, totaling approximately 7,771 hours of audio, with an average duration of around 4 minutes per song. The language distribution includes 49,692 Chinese songs and 66,797 English songs. Figure~\ref{fig:duration} shows the duration distributions for Chinese and English songs, indicating that both subsets exhibit similar length profiles suitable for long-form song modeling.

All samples consist of single-track audio with mixed vocals and accompaniment, paired with structured, time-aligned lyrics and hierarchical style annotations.

\section{Method}

Muse is designed to directly operate on the conversational, segment-structured data described in Section~\ref{sec:dataset}. By unifying text tokens and discrete audio tokens within a single sequence modeling framework, it supports long-form song generation with explicit segment-level style conditioning.

\subsection{Overview}
Muse is built upon a unified audio–language modeling framework that combines a Qwen-based language model with a neural audio codec (see Figure~\ref{intro}). Textual inputs, including global style labels, segment-level style descriptions, and lyrics, are tokenized using the standard Qwen tokenizer. Audio waveforms are encoded into discrete audio tokens using MuCodec \citep{xu2024mucodec}, an open-source Vector-Quantized Variational Autoencoder~\citep{DBLP:conf/nips/OordVK17} audio codec with a codebook size of 16,384.

The vocabulary of the language model is extended to incorporate the MuCodec audio tokens, enabling text and audio tokens to be modeled within a single autoregressive sequence. Given an input conversation consisting of textual prompts and previously generated audio tokens, the model predicts the next token using standard causal language modeling.

\subsection{Training}
We train Muse on the fully licensed synthetic dataset described in Section~\ref{sec:dataset}, which consists of 116,489 complete songs. Each song is represented as a multi-turn conversational sequence, where the first user message specifies global style attributes, and each subsequent turn provides a segment-level style description, lyrics, and phoneme information. The model predicts the corresponding span of audio tokens for each segment. Segment boundaries are predefined according to musical structure, including common sections such as Intro, Verse, Chorus, Bridge, and Outro.

Training is performed via standard supervised finetuning starting from Qwen3-0.6B \citep{DBLP:journals/corr/abs-2505-09388}, using a single-stage cross-entropy loss over the combined text and audio token vocabulary. We monitor the loss on a held-out validation set and select the checkpoint corresponding to the lowest validation loss to train for a total of 7 epochs. No diffusion models, reinforcement learning, auxiliary objectives, or task-specific losses are introduced. This minimal setup ensures that the entire system is fully reproducible using standard language model finetuning pipelines.

\subsection{Segment-Level Style Conditioning}
Fine-grained style control in Muse is achieved through a multi-turn conversational prompting scheme. Each song generation session begins with a user message specifying global style attributes, followed by a sequence of user messages that define the requirements for individual song segments. Each segment-level prompt includes a natural-language style description and the corresponding lyrics.

For each segment, the model generates a contiguous span of audio tokens as the assistant response. During training, the conversational data structure provides implicit supervision for segment boundaries through turn separation and segment-specific textual descriptions. Although we add [BOA] and [EOA] tokens to mark the start and end of each audio segment, no additional boundary detection modules or transition mechanisms are required. Under this training setup, we observe that the model is able to maintain coherence across segment boundaries, follow segment-level style descriptions, and preserve global stylistic consistency throughout the entire song.

\begin{table*}[t]
    \centering
    \caption{Main quantitative results for long-form song generation. We compare Muse with both closed-source systems (SUNO V4.5, SUNO V5, Mureka-O2) and open-source baselines (YuE, ACE-Step, LeVo, DiffRhythm~2) across phoneme error rate (PER), text--music similarity (MuLan-T), segment-level text--music similarity (MuLan-T$_{\text{seg}}$), audio aesthetic scores (CE, CU, PC, PQ), and SongEval metrics (CO, MU, ME, CL, NA).}
    \label{tab:main_results}
    \resizebox{\textwidth}{!}{%
    \begin{tabular}{lcccccccccccccc}
    \toprule
    \multirow{2}{*}{Model} & \multirow{2}{*}{Model Size} & \multirow{2}{*}{PER $\downarrow$} & \multirow{2}{*}{Mulan-T $\uparrow$} & \multirow{2}{*}{Mulan-T$_{\text{seg}}$ $\uparrow$} & \multicolumn{4}{c}{Audio Aesthetics $\uparrow$} & \multicolumn{5}{c}{SongEval $\uparrow$} \\
    \cmidrule(lr){6-9} \cmidrule(lr){10-14}
     &  &  &  &  & CE & CU & PC & PQ & CO & MU & ME & CL & NA \\
    \midrule
    \multicolumn{14}{c}{\textit{Closed-source Systems}} \\
    SUNO V4.5 & - & \textbf{0.09} & 0.39 & 0.34 & 7.67 & \textbf{7.84} & 6.30 & \textbf{8.32} & \textbf{4.60} & \textbf{4.51} & \textbf{4.60} & \textbf{4.54} & \textbf{4.46} \\
    SUNO V5 & - & 0.11 & \textbf{0.42} & \textbf{0.36} & \textbf{7.68} & 7.81 & 6.44 & 8.24 & \textbf{4.60} & 4.50 & 4.59 & 4.52 & 4.43 \\
    Mureka-O2 & - & \textbf{0.09} & 0.34 & - & 7.51 & 7.68 & \textbf{6.58} & 8.14 & 4.44 & 4.29 & 4.39 & 4.32 & 4.24 \\
    \midrule
    \multicolumn{14}{c}{\textit{Open-source Systems}} \\
    YuE & 8B & 0.37 & 0.27 & - & 7.10 & 7.58 & 5.72 & 7.98 & 3.41 & 3.14 & 3.25 & 3.17 & 3.10 \\
    ACE-Step & 3.5B & 0.39 & 0.28 & - & 6.83 & 7.13 & 6.19 & 7.31 & 3.39 & 3.15 & 3.20 & 3.21 & 3.08 \\
    LeVo & 5.1B & \textbf{0.14} & 0.26 & - & \textbf{7.61} & \textbf{7.84} & 6.22 & \textbf{8.32} & 3.58 & 3.42 & 3.43 & 3.46 & 3.35 \\
    DiffRhythm 2 & 1B & 0.15 & \textbf{0.37} & - & 7.36 & 7.51 & 5.60 & 8.08 & 3.48 & 3.30 & 3.36 & 3.34 & 3.18 \\
    Muse & 0.6B & 0.16 & 0.33 & \textbf{0.31} & 7.49 & 7.68 & \textbf{6.61} & 8.14 & \textbf{4.06} & \textbf{3.88} & \textbf{3.98} & \textbf{3.93} & \textbf{3.87} \\
    \bottomrule
    \end{tabular}%
    }
    \end{table*}
    
\begin{figure*}[t]
\centering
\begin{minipage}{0.48\textwidth}
\centering
\includegraphics[width=\textwidth]{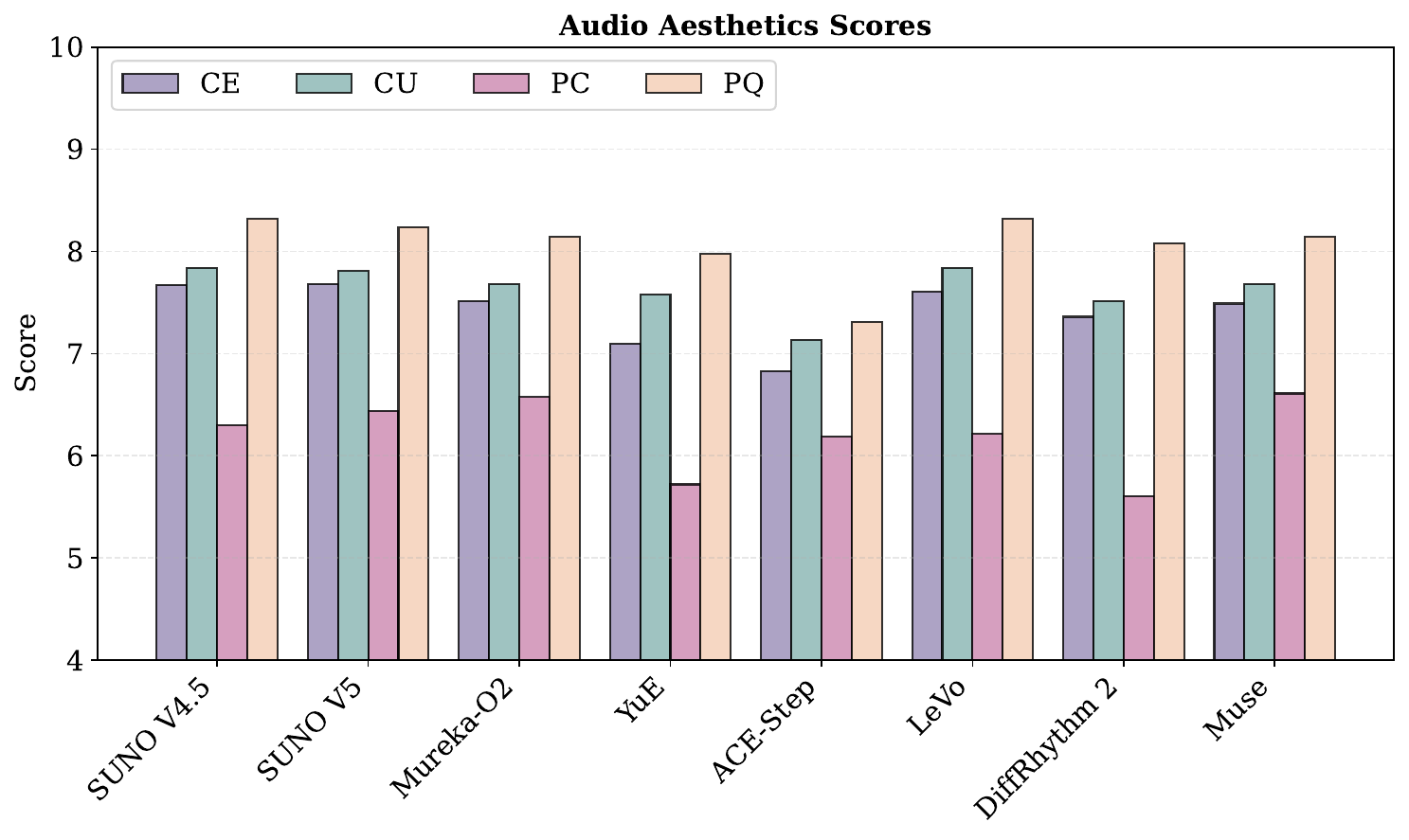}
\end{minipage}
\hfill
\begin{minipage}{0.48\textwidth}
\centering
\includegraphics[width=\textwidth]{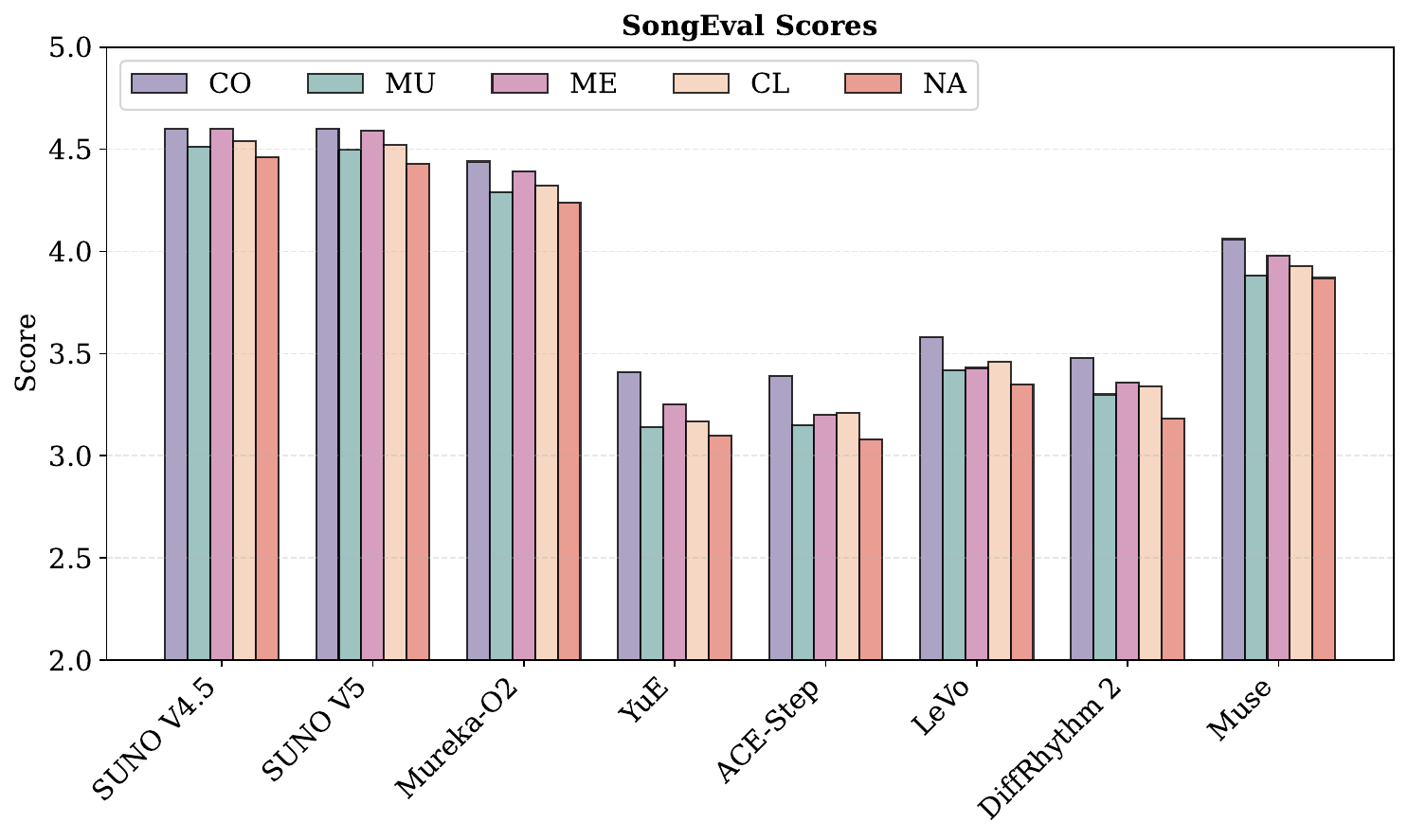}
\end{minipage}
\caption{Comparison of overall audio quality across models. Left: Meta Audiobox Aesthetics scores. Right: SongEval scores. Higher values indicate better performance.}
\label{fig:aesthetics}
\end{figure*}

\section{Experiments}
This section evaluates Muse on long-form song generation with a focus on generation quality, lyric alignment, and style controllability. We compare Muse with both open-source and closed-source systems under a unified evaluation protocol, and further analyze the impact of style annotations and model scaling.

\subsection{Experimental Setup}
\paragraph{Data Splits.}
From the full dataset described in Section~\ref{sec:dataset}, we reserve 256 songs for validation and 100 songs for testing. The remaining samples are used for training. All splits are disjoint at the song level to avoid any overlap in lyrics or audio content across splits.

\paragraph{Baselines.}
We compare Muse against a diverse set of existing systems. \textbf{Open-source baselines} include YuE, ACE-Step, LeVo, and Diffrhythm2, all of which release pretrained models and inference code but support only global style control. \textbf{Closed-source systems} include SunoV4.5  \citep{suno2024}, SunoV5, and Mureka-O2, which are evaluated using their official APIs. As these systems do not expose training data or model internals, we evaluate them solely based on their generated outputs, using the same automatic metrics as for open-source baselines.

\paragraph{Generation Protocol.}
All models are prompted to generate complete songs with structured lyrics. For systems that do not support segment-level style control, only global style prompts are provided. Muse is evaluated by default with both global style prompts and segment-level style descriptions, and variants using only global style prompting are considered only in ablation experiments.

\paragraph{Training Details.}
All models are trained for 7 epochs, with the final checkpoint selected based on the lowest validation loss. Training is performed on 8 NVIDIA H200 GPUs, taking approximately 150 minutes per epoch.

\begin{table*}[t]
    \centering
    \caption{Ablation study of Muse. Each row shows the effect of removing a specific component.}
    \label{tab:ablation}
    \resizebox{\textwidth}{!}{%
    \begin{tabular}{lcccccccccccc}
    \toprule
    \multirow{2}{*}{Model} & \multirow{2}{*}{PER $\downarrow$} & \multirow{2}{*}{Mulan-T $\uparrow$} & \multirow{2}{*}{Mulan-T$_{\text{seg}}$ $\uparrow$} & \multicolumn{4}{c}{Audio Aesthetics $\uparrow$} & \multicolumn{5}{c}{SongEval $\uparrow$} \\
    \cmidrule(lr){5-8} \cmidrule(lr){9-13}
     &  &  &  & CE & CU & PC & PQ & CO & MU & ME & CL & NA \\
    \midrule
    Muse & \underline{0.16} & \textbf{0.33} & \textbf{0.31} & \underline{7.49} & \underline{7.68} & 6.61 & 8.14 & \underline{4.06} & 3.88 & \underline{3.98} & \underline{3.93} & \underline{3.87} \\
    w/o phoneme & \textbf{0.14} & \textbf{0.33} & \underline{0.30} & 7.47 & 7.65 & 6.57 & 8.12 & \underline{4.06} & 3.88 & 3.95 & \underline{3.93} & 3.86 \\
    w/o segment style & \underline{0.16} & \textbf{0.33} & 0.28 & 7.47 & \textbf{7.69} & \textbf{6.66} & \underline{8.15} & \underline{4.09} & \underline{3.91} & \underline{4.00} & \underline{3.97} & \underline{3.91} \\
    w/o global style & 0.18 & \underline{0.32} & \underline{0.30} & \textbf{7.50} & \underline{7.68} & \underline{6.63} & \textbf{8.16} & \textbf{4.10} & \textbf{3.93} & \textbf{4.01} & \textbf{3.98} & \textbf{3.92} \\
    \bottomrule
    \end{tabular}%
    }
    \end{table*}

\subsection{Evaluation Metrics}
We adopt multiple complementary metrics to assess different aspects of song generation quality.

\paragraph{Phoneme Error Rate.}
We evaluate lyric-to-vocal alignment using phoneme error rate (PER). Specifically, we first transcribe the generated vocals into text using Qwen3-ASR \citep{qwen3asr2025}. Both the reference lyrics and the ASR transcriptions are converted into phoneme sequences using EmotiVoice \citep{emotivoice2024}. PER is computed as the normalized edit distance between the two phoneme sequences:
\begin{equation}
\mathrm{PER} = \frac{S + D + I}{N},
\end{equation}
where $S$, $D$, and $I$ denote the numbers of phoneme substitutions, deletions, and insertions, respectively, and $N$ is the total number of phonemes in the reference sequence. Lower PER indicates better alignment between the generated vocals and the target lyrics, and thus higher lyric fidelity. For songs whose length exceeds the model's maximum generation length, we truncate the reference lyrics to match the generated portion.

\paragraph{Text--Music Style Similarity.}
We evaluate the semantic alignment between textual style descriptions and generated audio using MuQ-MuLan. Let $E(\cdot)$ denote the embedding of a text or audio sequence. At the global-song level, the similarity between a global style description $s_g$ and the generated song $a$ is computed as cosine similarity:
\[
\text{Sim}_{\text{global}} = \frac{E(s_g) \cdot E(a)}{\|E(s_g)\| \, \|E(a)\|}.
\]
At the segment level, given $N$ segments in the song, each segment-level style description $s_i$ and corresponding audio segment $a_i$ are compared using cosine similarity, and the average across all segments is reported:
\[
\text{Sim}_{\text{segment}} = \frac{1}{N} \sum_{i=1}^{N} \frac{E(s_i) \cdot E(a_i)}{\|E(s_i)\| \, \|E(a_i)\|}.
\]
This evaluation captures both overall stylistic coherence and fine-grained adherence to segment-level style instructions.

\paragraph{Audio Aesthetic Quality.}
Overall audio quality is evaluated using automatic music aesthetic scoring models that correlate with human judgments of musicality and production quality. Specifically, we report scores from Meta Audiobox Aesthetics \citep{DBLP:journals/corr/abs-2502-05139} and SongEval \citep{yao2025songeval}. Meta Audiobox Aesthetics provides four metrics: content enjoyment (CE), content usefulness (CU), production complexity (PC), and production quality (PQ), which assess both subjective enjoyment and objective production characteristics. SongEval outputs five metrics: overall coherence (CO), memorability (ME), naturalness of vocal breathing and phrasing (NA), clarity of song structure (CL), and overall musicality (MU), capturing perceptual quality across structural, vocal, and holistic musical aspects.

\begin{table*}[t]
    \centering
    \caption{Model scaling results for Muse with fixed data and training setup.}
    \label{tab:model_scaling}
    \resizebox{\textwidth}{!}{%
    \begin{tabular}{lcccccccccccc}
    \toprule
    \multirow{2}{*}{Model Size} & \multirow{2}{*}{PER $\downarrow$} & \multirow{2}{*}{Mulan-T $\uparrow$} & \multirow{2}{*}{Mulan-T$_{\text{seg}}$ $\uparrow$} & \multicolumn{4}{c}{Audio Aesthetics $\uparrow$} & \multicolumn{5}{c}{SongEval $\uparrow$} \\
    \cmidrule(lr){5-8} \cmidrule(lr){9-13}
     &  &  &  & CE & CU & PC & PQ & CO & MU & ME & CL & NA \\
    \midrule
    Muse (0.6B) & 0.16 & 0.33 & \underline{0.31} & \underline{7.49} & \textbf{7.68} & \underline{6.61} & \underline{8.14} & \underline{4.06} & 3.88 & \underline{3.98} & 3.93 & 3.87 \\
    1.7B & 0.18 & 0.32 & \underline{0.31} & 7.28 & \underline{7.67} & 6.32 & 8.12 & 3.91 & 3.74 & 3.83 & 3.78 & 3.73 \\
    4B & \underline{0.14} & \underline{0.34} & \underline{0.31} & \textbf{7.53} & \underline{7.67} & \textbf{6.66} & \textbf{8.15} & \textbf{4.09} & \textbf{3.92} & \textbf{3.99} & \textbf{3.97} & \textbf{3.91} \\
    8B & \textbf{0.12} & \textbf{0.35} & \textbf{0.32} & \underline{7.49} & 7.66 & \underline{6.61} & 8.12 & \underline{4.06} & \underline{3.89} & \underline{3.97} & \underline{3.94} & \underline{3.89} \\
    \bottomrule
    \end{tabular}%
    }
    \end{table*}

\subsection{Main Results}

Table~\ref{tab:main_results} summarizes the main quantitative results comparing Muse with both open-source and closed-source song generation systems.

\paragraph{Overall Performance.}
Muse achieves strong and well-balanced performance among open-source models across all evaluated metrics. Despite its small model size (0.6B parameters), Muse consistently outperforms prior open systems on most Audio Aesthetics and SongEval metrics (see Figure~\ref{fig:aesthetics}), while remaining competitive on lyric alignment and global style similarity. These results suggest that effective supervision and data design can compensate for model scale in long-form song generation.

\paragraph{Lyric Alignment and Style Similarity.}
Muse achieves substantially lower phoneme error rates than most open-source baselines, indicating improved alignment between generated vocals and lyrics. On global style similarity measured by Mulan-T, Muse performs comparably to DiffRhythm~2 while maintaining stronger perceptual quality and musical coherence, as reflected by Audio Aesthetics and SongEval scores.

\paragraph{Segment-Level Style Control.}
Muse is able to follow fine-grained segment-level style descriptions, achieving a Mulan-T similarity of 0.31 across song segments. For comparison, SunoV5 and SunoV4.5 reach 0.36 and 0.34, respectively. This demonstrates that Muse captures segment-specific stylistic nuances effectively while maintaining global coherence.

\paragraph{Comparison with Closed-Source Systems.}
Compared with commercial systems such as Suno and Mureka, Muse narrows the performance gap on both style similarity and audio quality, although closed-source models still achieve higher absolute aesthetic scores. Notably, Muse attains these results using a fully open and reproducible training pipeline, without access to proprietary data or training procedures.

\paragraph{Decoding and Reproducibility.}
To ensure reproducibility, all models are evaluated using deterministic or controlled decoding by fixing random seeds or setting the sampling temperature to zero whenever feasible. We observe that several open-source models, including YuE, ACE-Step, LeVo, and Muse, exhibit degraded generation quality under strictly deterministic decoding, often producing repetitive outputs. In particular, LeVo fails to generate valid samples with temperature set to zero and is therefore evaluated with a temperature of 0.9. Similarly, Muse encounters rare generation failures under zero-temperature decoding for a small fraction of samples; these cases are regenerated using a temperature of 0.9. Detailed statistics and sample-level information for these cases are reported in the appendix, along with full evaluation results obtained by decoding the entire test set with a temperature of 0.9. This behavior reflects a broader challenge in reproducible evaluation of autoregressive music generation models.

Overall, these results demonstrate that Muse provides a competitive and fully reproducible baseline for long-form song generation across lyric alignment, style controllability, and musical quality.

\subsection{Ablation and Scaling Studies}
\paragraph{Component Ablation.}
We evaluate variants of Muse with individual components removed, including phoneme supervision, segment-level style descriptions, and global style prompts (Table~\ref{tab:ablation}). Removing phoneme supervision slightly improves PER, which we attribute to the phoneme annotations providing low-level acoustic information without semantic content, potentially introducing redundancy during training. Omitting segment-level style reduces Mulan-T\_seg, confirming its role in guiding local stylistic transitions, while removing global style decreases overall Mulan-T and audio aesthetic scores, highlighting its importance for coherent song-level expression. These results demonstrate that each component contributes complementary benefits for lyric fidelity, segment-level style controllability, and overall musical quality.

\paragraph{Model Scaling.}
We study the effect of model scale by training Muse with language models ranging from 0.6B to 8B parameters, keeping the training data and procedure fixed. As shown in Table~\ref{tab:model_scaling}, increasing model size generally improves lyric alignment and global style similarity, while segment-level style controllability remains largely stable around 0.31--0.32. The phoneme error rate decreases with scale, reaching its lowest value for the 8B model.

Larger models also achieve slight gains in musical coherence and perceptual quality, reflected in higher SongEval scores on several dimensions. Audio aesthetic improvements are modest beyond mid-scale, indicating diminishing returns in perceived quality. Overall, these results suggest that Muse benefits from scaling, particularly for lyric fidelity and structural control, while maintaining strong performance even at the 0.6B scale.

\section{Conclusion}
We introduce a fully open-source system for long-form song generation with fine-grained style conditioning, comprising a licensed synthetic dataset, complete training and evaluation pipelines, and Muse, an easy-to-deploy song generation model. By releasing all components required for reproduction, this work addresses a major reproducibility challenge in academic song generation research.

Muse adopts a unified audio--language modeling framework and is trained via single-stage supervised finetuning without task-specific losses. Despite its modest model size and data scale, Muse achieves competitive performance across lyric alignment, style similarity at both global and segment levels, and audio aesthetic quality, while supporting explicit control over musical structure.

Beyond model performance, our results demonstrate that data organization and supervision design play a crucial role in controllable long-form music generation. This work provides a reproducible and extensible foundation for future research on structure-aware and data-efficient song generation, enabling fair comparison and analysis.
\section{Limitations}
While Muse demonstrates strong performance on controllable long-form song generation, several limitations remain.

First, Muse relies on synthetic training data generated by existing commercial systems. Although all data are fully licensed and enable reproducible research, the distribution of the dataset may inherit biases from the underlying generation models, potentially limiting stylistic diversity.

Second, Muse's generation quality can degrade as song length increases. While the model is trained on full-length songs, extremely long compositions may accumulate errors in melody, rhythm, or style consistency, which could affect overall coherence and musicality.

Third, evaluation relies primarily on automatic metrics for lyric alignment, style similarity, and audio aesthetics. Although these metrics correlate with human judgment, they do not fully capture subjective musical qualities such as emotional expressiveness or creative originality.

Finally, Muse does not aim to perform voice cloning or artist-specific style imitation. While this design choice avoids ethical and legal risks, it also limits the model's ability to generate songs in highly specific vocal identities.

Addressing these limitations, particularly through more diverse licensed data and richer human evaluation, remains an important direction for future work.

\bibliography{main}


\appendix
\onecolumn
\appendix

\section{AI Assistants in Research or Writing}

In preparing this manuscript, AI assistants were employed solely to assist with refining the clarity, style, and readability of certain text segments. They were not involved in designing the study, developing or implementing the methodology, collecting or analyzing data, or generating the primary scientific contributions. All substantive research decisions, analyses, and conclusions are fully the responsibility of the authors.

\section{Data Generation}

Training data are generated via a two-stage pipeline. GPT is first used to produce structured lyrics and music style descriptions. These textual prompts are then fed into the Suno model to generate paired songs and metadata, which are collected as the final training dataset.

\paragraph{GPT-based Data Generation Prompts.}

We provide the system and user prompts used to guide the GPT in generating structured training data, including lyrics and high-level musical style descriptions.

\begin{tcolorbox}[
    colback=white,
    colframe=gray!70!black,
    title=System Prompt,
    coltitle=white,
    fonttitle=\bfseries,
    center title,
    rounded corners,
    boxrule=0.6mm,
    width=\textwidth,
    breakable,
    enhanced,
    left=6pt,
    right=6pt,
    top=4pt,
    bottom=4pt
]

\begin{Verbatim}[breaklines]
You are a creative music lyricist and composer. Please generate diverse and creative music tag-based descriptions and LRC format lyrics with song structure tags. CRITICAL REQUIREMENTS: 1) Description must be structured tags separated by commas, NOT narrative text. 2) Return ONLY pure, valid JSON format without any extra symbols, markers, or comments. 3) Each song must include structure tags like [Verse 1], [Chorus], [Bridge], etc., followed by LRC format lyrics [mm:ss.xx]lyric_content. 4) MANDATORY: Each song must have MORE than {require_length} lines of lyrics with timestamps. 
\end{Verbatim}

\vspace{0.5em}

\end{tcolorbox}

\begin{tcolorbox}[
    colback=white,
    colframe=gray!70!black,
    title=User Prompt,
    coltitle=white,
    fonttitle=\bfseries,
    center title,
    rounded corners,
    boxrule=0.6mm,
    width=\textwidth,
    breakable,
    enhanced,
    left=6pt,
    right=6pt,
    top=4pt,
    bottom=4pt
]

\begin{Verbatim}[breaklines]
Generate 2 complete songs. Each song must meet the following hard requirements:

- Strictly forbidden to generate lyrics with fewer than {require_length} lines!
- The number of lyric lines for each song must be strictly greater than {require_length}. This is a hard requirement!
- The timestamp of the final line must be between {start_timestamp} and {end_timestamp}.
- The two songs must differ in duration and line count; their final timestamps must not be identical.
- The timestamp interval between adjacent lyric lines must not exceed 10 seconds! Timestamps must be continuous and progress naturally.
- Awkward gaps like "[03:25.00]in the heart[04:25.00]the last lyric" are strictly forbidden. Do not exceed a 10-second interval.
- It is strictly forbidden to repeat the entire structure or its sections after one iteration is complete. It is also strictly forbidden to repeat the same lyric line multiple times.
If any of the above requirements are not met, the generation is considered a failure. Please regenerate.
Please generate 2 new, diverse music descriptions and LRC format lyrics. The language should be English.

Creative Requirements:

1. Style and Genre must be diverse.
2. Description Tagging Requirements (Must be strictly followed):
   The description field must use a structured tag format, including the following tags, separated by commas:
   - Music Style tag
   - Music Genre tag
   - Instruments tag
   - Emotional Tone tag
   - Mood/Atmosphere tag
   - Vocal Style and Voice tag, limited to either "male voice" or "female voice", solo performance only.
   Note: Each tag should be concise. Multiple tags of the same category can be separated by a slash (e.g., "Piano/Violin").
3. Lyric Creativity: The lyrics should have depth and artistry:
   - Themes can cover various aspects such as love, life, society, nature, philosophy, dreams, memories, etc.
   - Use rich literary devices: metaphors, imagery, contrast, parallelism, etc.
   - Express sincere emotions with a focus on rhyme and rhythm.
   - The style can be narrative, lyrical, or stream-of-consciousness.
4. Lyric Structure and Length Requirements (Must be strictly followed):
   - The lyrics must be organized using the following structure, with section tags annotating each part.
   - The structure must strictly follow this order, for a total of 8 section tags: [Verse 1] → [Pre-Chorus] → [Chorus] → [Verse 2] → [Pre-Chorus] → [Chorus] → [Bridge] → [Chorus (Outro)].
   - A single song can only have these 8 section tags. [Verse 1] and [Verse 2] appear once; [Pre-Chorus] and [Chorus] appear twice; [Bridge] and [Chorus (Outro)] appear once. Do not add or repeat extra section tags.
   - Each section tag (e.g., [Verse 1], [Chorus]) must be on its own line, immediately followed by the LRC format lyrics for that section.
   - Separate sections with a blank line.
   - **Total Line Count Requirement**: The entire song must contain at least {require_length} lines of timestamped lyrics (not including section tags or blank lines).
5. LRC Format Mandatory Rules (Must be strictly followed):
   - Each line of lyrics must be in the format `[mm:ss.xx]Lyric content`, with no space between the timestamp and the lyrics. The lyric content should be coherent.
   - **Each line must contain only one short phrase of lyrics.** Start a new line when encountering punctuation like commas or periods.
   - **Strictly forbidden to merge multiple sentences or clauses onto the same line.**
   - Timestamps must be distributed naturally. **The first line's timestamp must not be [00:00.00]**. Allow for an instrumental intro (suggestion: start between [00:05.00] and [00:15.00]).
   - Timestamp intervals must be varied: The intervals within each song must be diverse, often using decimal values. Do not use a fixed interval:
     * A single song must contain a variety of different intervals; do not use the same interval for all lines (e.g., not all 4-second gaps).
     * Dynamically adjust intervals based on the emotional intensity and rhythm of the lyrics.
     * The gap between adjacent lines should vary to reflect the musical rhythm.
   - Timestamp allocation should be reasonably inferred based on the song's style, emotion, and rhythm, not mechanically assigned based on lyric length.
   - The length of each lyric line should vary naturally; do not make them all uniform.
   - **The total song duration must be between {start_duration} and {end_duration} (meaning the final line's timestamp must be between {start_timestamp} and {end_timestamp}). This is a hard requirement!**
6. Lyric Length Requirement: The number of lyric lines in the lyrics field must be greater than {require_length}. If the generated length is too short, please regenerate.
7. Uniqueness and Originality: Each piece should be unique. Avoid simply repeating the content from examples.
8. Format Requirements:
   - Directly return a JSON array containing 2 song objects. Each object must have only "description" and "lyrics" fields.
   - `description` field: Must be in tag format, not narrative text.
   - `lyrics` field: A string in LRC format with section tags.
   - Strictly forbidden to insert any extra symbols, markers, comments, or explanatory text within the JSON.

LRC Format Example (with section tags):
[Verse 1]
[00:08.00]First line of lyrics
[00:12.50]Second line of lyrics
[00:17.20]Third line of lyrics

[Pre-Chorus]
[00:22.00]Pre-chorus lyrics
[00:26.50]Pre-chorus lyrics

[Chorus]
[00:31.00]Chorus lyrics
[00:35.50]Chorus lyrics

Negative Examples (to avoid):
- Incorrect: [01:30.00](Piano Interlude) - Do not add parenthetical comments after the timestamp.
- Incorrect: [00:00.00]Starting lyric - The first line cannot start at 00:00.00.
- Incorrect: [00:05.00]In the familiar field, the sun casts golden rays upon the wheat - Strictly forbidden to place multiple clauses on the same line.
- Incorrect: [03:00.00] In the light of hope[03:05.50] In the light of hope[03:10.20] In the light of hope -Excessive repetition of the exact same lyric line is strictly forbidden. Lyrical content must show variation.
Now, please fully unleash your creativity and generate 2 new, complete works of music descriptions and LRC format lyrics.
Special Reminder: Each song must be complete, not abbreviated or omitted! It must contain the full 8 sections (Verse 1, Pre-Chorus, Chorus, Verse 2, Pre-Chorus, Chorus, Bridge, Chorus Outro) and strictly ensure more than {require_length} lines of lyrics.

Directly return in JSON array format:
[
  {{"description": "...", "lyrics": "..."}},
  {{"description": "...", "lyrics": "..."}}
]
\end{Verbatim}

\vspace{0.5em}

\end{tcolorbox}

\paragraph{Suno Model Response Example.}

An example of the metadata returned by the Suno music generation model is shown to illustrate the intermediate outputs used in the data pipeline, such as style tags, duration, and generation conditions.

\begin{tcolorbox}[
    colback=white,
    colframe=gray!70!black,
    title=Suno Info,
    coltitle=white,
    fonttitle=\bfseries,
    center title,
    rounded corners,
    boxrule=0.6mm,
    width=\textwidth,
    breakable,
    enhanced,
    left=6pt,
    right=6pt,
    top=4pt,
    bottom=4pt
]

\begin{Verbatim}[breaklines]
{
  "song_id": "sunov5_000001",
  "song_index": 1,
  "track_index": 0,
  "lyrics": "[Verse 1: The song begins with a gentle, melancholic piano melody, ...]\nIn the morning,\nI see you...[Pre-Chorus1: ...]\nWe chase together,\nThe shape of Dream\nLeave footprint in the sky...",
  "timestamped_lyrics": {
    "alignedWords": [
      {
        "word": "[Verse 1: The song begins with a gentle, melancholic piano melody...]\nIn the morning,\n",
        "success": true,
        "startS": 13.64362,
        "endS": 15.67819,
        "palign": 0
      },
      {
        "word": "I see you\n",
        "success": true,
        "startS": 15.79787,
        "endS": 17.15426,
        "palign": 0
      },
      {
        "...": "..."
      },
      {
        "word": "[Pre-Chorus1: ...]\nWe chase together,\n",
        "success": true,
        "startS": 34.22998,
        "endS": 36.02394,
        "palign": 0
      },
      {
        "word": "The shape of Dream\n",
        "success": true,
        "startS": 36.14362,
        "endS": 37.57979,
        "palign": 0
      },
      {
        "word": "Leave footprint in the sky\n\n",
        "success": true,
        "startS": 37.73936,
        "endS": 42.92616,
        "palign": 0
      },
      {
        "...": "..."
      }
    ],
    "waveformData": [
      8e-05,
      4e-05,
      0.01169,
      0.01437,
      0.01525,
      0.03603
    ],
    "hootCer": 0.9214705242549489,
    "isStreamed": false
  },
  "style": "Pop, Ballad, C-pop, Romantic, Soft Rock, Piano, Strings, Electric Guitar, Female Vocal.",
  "full_track_data": {
    "id": "f8fd4420-43c1-4fee-a688-0370a7185895",
    "audioUrl": "https://musicfile.api.box/xxx.mp3",
    "sourceAudioUrl": "https://cdn1.suno.ai/xxx.mp3",
    "streamAudioUrl": "https://musicfile.api.box/xxx",
    "sourceStreamAudioUrl": "https://cdn1.suno.ai/xxx",
    "imageUrl": "https://musicfile.api.box/xxx.jpeg",
    "sourceImageUrl": "https://cdn2.suno.ai/image_xxx.jpeg",
    "prompt": "prompt",
    "modelName": "chirp-crow",
    "title": "Song_sunov5_000001",
    "tags": "Pop, Ballad, C-pop, Romantic, Soft Rock, Piano, Strings, Electric Guitar, Female Vocal.",
    "createTime": 1766820274832,
    "duration": 234.12
  }
}
\end{Verbatim}

\vspace{0.5em}

\end{tcolorbox}

\section{Supplementary Training Details}

\paragraph{Training Data Sample.}

After paragraph re-annotation, phoneme extraction, and other processing operations, we finally organized the training data into the following format.

\begin{tcolorbox}[
    colback=white,
    colframe=gray!70!black,
    title=Training Message,
    coltitle=white,
    fonttitle=\bfseries,
    center title,
    rounded corners,
    boxrule=0.6mm,
    width=\textwidth,
    breakable,
    enhanced,
    left=6pt,
    right=6pt,
    top=4pt,
    bottom=4pt
]

\begin{Verbatim}[breaklines]
[
  {
    "messages": [
      {
        "role": "user",
        "content": "Please generate a song in the following style:Pop, Ballad, C-pop, Romantic, Soft Rock, Piano, Strings, Electric Guitar, Female Vocal.\nNext, I will tell you the requirements and lyrics for the song fragment to be generated, section by section.\n[Intro][desc:The track opens with a gentle...]"
      },
      {
        "role": "assistant",
        "content": "[SOA]<AUDIO_7224><AUDIO_5151><AUDIO_15457>...[EOA]"
      },
      {
        "role": "user",
        "content": "[Verse 1][desc:The song begins with a melancholic piano melody...][lyrics:\nIn the sky,\nI see you\n...][phoneme:\nIH0 N DH AH0S K AY1\nAY1 S IY1 Y UW1\n]"
      },
      {
        "role": "assistant",
        "content": "[SOA]<AUDIO_12107><AUDIO_5505><AUDIO_15590>...[EOA]"
      },
      {
        "role": "user",
        "content": "[Pre-Chorus1][desc:...][lyrics:\n...][phoneme:\n...]"
      },
      {
        "role": "assistant",
        "content": "[SOA]<AUDIO_5911><AUDIO_2317><AUDIO_5114>...[EOA]"
      },
      {
        "...": "..."
      }
    ]
  }
]
\end{Verbatim}

\vspace{0.5em}

\end{tcolorbox}

\paragraph{Training Script.}
The following script illustrates the distributed training configuration used for model optimization, including environment setup, multi-GPU initialization, and key training hyperparameters. Training is performed using full-parameter fine-tuning with DeepSpeed ZeRO-3 for memory-efficient large-scale optimization.

\begin{tcolorbox}[
    colback=white,
    colframe=gray!70!black,
    title=Training Script,
    coltitle=white,
    fonttitle=\bfseries,
    center title,
    rounded corners,
    boxrule=0.6mm,
    width=\textwidth,
    breakable,
    enhanced,
    left=6pt,
    right=6pt,
    top=4pt,
    bottom=4pt
]

\begin{Verbatim}[breaklines]
#!/usr/bin/env bash
source /root/miniconda3/etc/profile.d/conda.sh
conda activate <conda_env>

export NCCL_DEBUG=WARN

export ARNOLD_WORKER_GPU=8
export ARNOLD_WORKER_NUM=1
export ARNOLD_ID=0
export ARNOLD_WORKER_0_HOST=127.0.0.1
export ARNOLD_WORKER_0_PORT=29500

export NPROC_PER_NODE=$ARNOLD_WORKER_GPU
export MASTER_PORT=${ARNOLD_WORKER_0_PORT:-29500}
export NNODES=$ARNOLD_WORKER_NUM
export NODE_RANK=$ARNOLD_ID
export MASTER_ADDR=$ARNOLD_WORKER_0_HOST
export LOCAL_WORLD_SIZE=$ARNOLD_WORKER_GPU
export WORLD_SIZE=$((ARNOLD_WORKER_NUM * ARNOLD_WORKER_GPU))

export RUN_NAME="Muse_0.6b_main_5e-4"
MODEL_PATH="qwen3-0.6B-music"
OUTPUT_DIR="${RUN_NAME}"

if [ $NODE_RANK -eq 0 ]; then
    mkdir -p ${OUTPUT_DIR}
    echo "Starting multi-node training with $NNODES nodes, $NPROC_PER_NODE GPUs each"
    echo "Total GPUs: $WORLD_SIZE"
fi

sleep 5

swift sft \
    --model ${MODEL_PATH} \
    --train_type full \
    --model_type qwen3 \
    --dataset 'whole_train_cn_1.jsonl'\
             'whole_train_en_1.jsonl' \
    --val_dataset 'whole_valid_1.jsonl' \
    --num_train_epochs 20 \
    --learning_rate 5e-4 \
    --per_device_train_batch_size 1 \
    --gradient_accumulation_steps 8 \
    --save_steps -1 \
    --save_strategy epoch \
    --eval_strategy epoch \
    --save_total_limit 200 \
    --save_only_model true \
    --logging_steps 1 \
    --max_length 15000 \
    --output_dir ${OUTPUT_DIR} \
    --warmup_ratio 0.05 \
    --dataloader_num_workers 32 \
    --dataset_num_proc 8 \
    --deepspeed zero3 \
    --report_to tensorboard \
    2>&1 | tee ${OUTPUT_DIR}/train_node_${NODE_RANK}.log
\end{Verbatim}

\vspace{0.5em}

\end{tcolorbox}

\section{Supplementary Analysis on Decoding Stability and Evaluation Robustness}
\label{appendix:decoding_stability}

\subsection{Deterministic Decoding for Reproducibility}

To ensure strict experimental rigor and full reproducibility, all main evaluations in this paper adopt deterministic decoding by setting the temperature to $T=0$ during inference.
Under this setting, identical inputs deterministically produce identical outputs across different runs and checkpoints, which is essential for fair comparison and reliable error attribution.

However, in the context of modality-extended generation, deterministic decoding may introduce inference-time instability.
In particular, for certain checkpoints and evaluation samples, the model may fail to produce valid outputs, manifesting as immediate token-level repetition or early decoding collapse.
Importantly, these failures do not consistently occur on specific subsets of the evaluation data, nor are they associated with particular semantic categories or difficulty levels.

\subsection{Failure Analysis and Interpretation}

Initially, one plausible hypothesis was that these failures arose from intrinsically difficult evaluation samples.
However, qualitative inspection and statistical analysis reveal that failed generations are sparsely and randomly distributed across the test set.
Different checkpoints exhibit failures on different samples, and no systematic correlation with input length, content domain, or perceived complexity is observed.

This evidence suggests that such failures should not be attributed to data difficulty or curriculum effects.
Instead, we interpret them as \emph{decoding instabilities induced by deterministic inference} under modality extension.
Most models evaluated in this work are trained via supervised fine-tuning (SFT) on carefully curated multimodal data, without additional post-training stages aimed at robustness enhancement.
As a result, while the models are capable of high-quality generation, their inference-time stability under strict greedy decoding may be limited.

From an evaluation perspective, these failures are therefore best regarded as \emph{exogenous variables}—artifacts of the decoding strategy rather than reflections of the model’s underlying capability.

\subsection{Accounting for Failed Generations in Evaluation}

A critical challenge then arises: how should failed generations be incorporated into quantitative evaluation?
Simply discarding failed samples reduces the effective test set size and introduces optimistic bias, as unfavorable cases are selectively removed.
Conversely, assigning arbitrary low scores to failed outputs would inject artificial noise and disproportionately penalize certain checkpoints.

To preserve test set completeness while avoiding biased scoring, we adopt the following protocol.
All generations are first attempted using deterministic decoding ($T=0$).
For samples where decoding fails, we re-generate the corresponding outputs using stochastic decoding with $T=0.9$ and Top-$p=0.9$.
The resulting generations are then merged back into the original evaluation set, ensuring that each test sample contributes exactly one valid output to the final metrics.

This strategy treats deterministic decoding failures as external inference-time events and avoids conflating them with model performance.
The proportion of samples requiring stochastic re-sampling (\texttt{do\_sample}) is explicitly reported (6.7\% of all evaluation samples) to ensure transparency in the evaluation protocol.

\subsection{Empirical Validation of the Evaluation Protocol}

To validate the rationality of this protocol, we conduct a supplementary comparison using two alternative evaluation settings:
(1) deterministic decoding at $T=0$ with all failed samples removed, and
(2) fully stochastic decoding with $T=0.9$ applied to all samples.

The quantitative results are shown in Table~\ref{tab:t0_vs_t09}.

\begin{table*}[t]
    \centering
    \caption{Comparison of Model Performance: Deterministic ($T=0$) vs. Stochastic ($T=0.9$)} 
    \label{tab:t0_vs_t09}
    \resizebox{\textwidth}{!}{%
    \begin{tabular}{lcccccccccccc}
    \toprule
    \multirow{2}{*}{Model} & \multirow{2}{*}{PER $\downarrow$} & \multirow{2}{*}{Mulan-T $\uparrow$} & \multirow{2}{*}{Mulan-T$_{\text{seg}}$ $\uparrow$} & \multicolumn{4}{c}{Audio Aesthetics $\uparrow$} & \multicolumn{5}{c}{SongEval $\uparrow$} \\
    \cmidrule(lr){5-8} \cmidrule(lr){9-13}
     &  &  &  & CE & CU & PC & PQ & CO & MU & ME & CL & NA \\
    \midrule
    Muse-0.6b & 0.16 & 0.33 & 0.31 & 7.49 & 7.68 & 6.61 & 8.14 & 4.06 & 3.88 & 3.98 & 3.93 & 3.87  \\
    Muse-0.6b-T0.9 & 0.17 & 0.34 & 0.31 & 7.47 & 7.62 & 6.76 & 8.10 & 4.02 & 3.84 & 3.94 & 3.90 & 3.84 \\
    \bottomrule
    \end{tabular}%
    }

    \end{table*}

We observe that evaluation scores under full stochastic decoding differ only slightly from those obtained by deterministic decoding with failed samples removed, with no method showing clear superiority. Our proposed protocol preserves the full test set while yielding performance estimates that are nearly identical, ensuring a complete and unbiased evaluation.

Together with the reported \texttt{do\_sample} ratio, these results support the validity of our approach for handling decoding failures without distorting evaluation outcomes.

\subsection{Summary}

In summary, deterministic decoding is adopted in this work to ensure reproducibility, but it may induce rare inference-time failures in modality-extended generation.
These failures are not data-dependent and should be treated as exogenous to model capability.
By selectively re-sampling failed generations and explicitly reporting their proportion, we maintain evaluation completeness while avoiding optimistic bias.
This appendix clarifies that the reported results reflect a principled balance between experimental rigor and robust performance estimation.

\end{document}